\documentclass{article}
\usepackage{graphicx} 
\usepackage{algorithmic}
\usepackage{algorithm}
\usepackage{amsmath}
\usepackage{tikz}
\usepackage{url} 
\usepackage[english]{babel}
\usepackage{natbib}
\setcitestyle{square}
\title{Robust estimation of causal dose-response relationship using exposure data with dose as an instrumental variable}
\author{Jixian Wang$^1$, Zhiwei Zhang$^2$, Ram Tiwari$^3$\\
$^1$Bristol Myers Squibb, Switzerland\\
$^2$Gilead Sciences, USA\\
$^3$Bristol Myers Squibb, USA}
\date{}

\begin{document}
\maketitle
\newcommand{\refb}[1]{(\ref{#1})}
\newcommand{\rb}{\right]}
\newcommand{\lb}{\left[}
\newcommand{\lpp}{\left(}
\newcommand{\rpp}{\right)}
\newcommand{\balpha}{{\mbox {\boldmath$\alpha$}}}
\newcommand{\hbalpha}{{\hat {{\mbox {\boldmath$\alpha$}}}}}
\newcommand{\halpha}{{\hat{\alpha}}}
\newcommand{\bbeta}{{\mbox {\boldmath$\beta$}}}
\newcommand{\hbbeta}{{\hat {\mbox {\boldmath$\beta$}}}}
\newcommand{\bdelta}{{\mbox {\boldmath$\delta$}}}
\newcommand{\bDelta}{{\mbox {\boldmath$\Delta$}}}
\newcommand{\bgamma}{{\mbox {\boldmath$\gamma$}}}
\newcommand{\bGamma}{{\mbox {\boldmath$\Gamma$}}}
\newcommand{\blambda}{{\mbox {\boldmath$\lambda$}}}
\newcommand{\bLambda}{{\mbox {\boldmath$\Lambda$}}}
\newcommand{\bdgm}{{\mbox {\boldmath$\digamma$}}}
\newcommand{\bPsi}{{\mbox {\boldmath$\Psi$}}}
\newcommand{\bpsi}{{\mbox {\boldmath$\psi$}}}
\newcommand{\bchi}{{\mbox {\boldmath$\chi$}}}
\newcommand{\bpi}{{\mbox {\boldmath$\pi$}}}
\newcommand{\bphi}{{\mbox {\boldmath$\phi$}}}
\newcommand{\bPhi}{{\mbox {\boldmath$\Phi$}}}
\newcommand{\btheta}{{\mbox {\boldmath$\theta$}}}
\newcommand{\bTheta}{{\mbox {\boldmath$\Theta$}}}
\newcommand{\btau}{{\mbox {\boldmath$\tau$}}}
\newcommand{\bepsi}{{\mbox {\boldmath$\varepsilon$}}}
\newcommand{\hbepsi}{ {\hat {\mbox {\boldmath$\varepsilon$}}}}
\newcommand{\epsi}{\varepsilon}
\newcommand{\bmu}{{\mbox {\boldmath$\mu$}}}
\newcommand{\bnu}{{\mbox {\boldmath$\nu$}}}
\newcommand{\beeta}{{\mbox {\boldmath$\eta$}}}
\newcommand{\bomega}{{\mbox {\boldmath$\omega$}}}
\newcommand{\bzeta}{{\mbox {\boldmath$\zeta$}}}
\newcommand{\bsigma}{{\mbox {\boldmath$\sigma$}}}
\newcommand{\bSigma}{{\mbox {\boldmath$\Sigma$}}}
\newcommand{\hsi}{{\hat{\sigma}}}
\newcommand{\hSi}{{\hat{\Sigma}}}
\newcommand{\bOmega}{{\mbox {\boldmath$\Omega$}}}
\newcommand{\bxi}{{\mbox {\boldmath$\xi$}}}
\newcommand{\bXi}{{\mbox {\boldmath$\Xi$}}}
\newcommand{\tra}{^{\top}}
\newcommand{\bv}{\mbox {\bf v}}
\newcommand{\bV}{\mbox {\bf V}}
\newcommand{\bX}{\mbox {\bf X}}
\newcommand{\bZ}{\mbox {\bf Z}}
\newcommand{\ba}{\mbox {\bf a}}
\newcommand{\ble}{\mbox {\bf e}}
\newcommand{\ha}{\hat{ a}}
\newcommand{\hba}{\hat{{\mbox {\bf a}}}}
\newcommand{\bc}{\mbox {\bf c}}
\newcommand{\bA}{\mbox {\bf A}}
\newcommand{\bB}{\mbox {\bf B}}
\newcommand{\bC}{\mbox {\bf C}}
\newcommand{\bD}{\mbox {\bf D}}
\newcommand{\bE}{\mbox {\bf E}}
\newcommand{\bF}{\mbox {\bf F}}
\newcommand{\mathF}{\mathcal{F}}
\newcommand{\bG}{\mbox {\bf G}}
\newcommand{\bg}{\mbox {\bf g}}
\newcommand{\bH}{\mbox {\bf H}}
\newcommand{\bI}{\mbox {\bf I}}
\newcommand{\bJ}{\mbox {\bf J}}
\newcommand{\bL}{\mbox {\bf L}}
\newcommand{\bM}{\mbox {\bf M}}
\newcommand{\bN}{\mbox {\bf N}}
\newcommand{\bp}{\mbox {\bf p}}
\newcommand{\bO}{\mbox {\bf O}}
\newcommand{\bP}{\mbox {\bf P}}
\newcommand{\bQ}{\mbox {\bf Q}}
\newcommand{\bK}{\mbox {\bf K}}
\newcommand{\bR}{\mbox {\bf R}}
\newcommand{\bT}{\mbox {\bf T}}
\newcommand{\bU}{\mbox {\bf U}}
\newcommand{\bS}{\mbox {\bf S}}
\newcommand{\bW}{\mbox {\bf W}}
\newcommand{\bY}{\mbox {\bf Y}}
\newcommand{\bb}{\mbox {\bf b}}
\newcommand{\bd}{\mbox {\bf d}}
\newcommand{\be}{\mbox {\bf e}}
\newcommand{\blf}{\mbox {\bf f}}
\newcommand{\bk}{\mbox {\bf k}}
\newcommand{\bh}{\mbox {\bf h}}
\newcommand{\bm}{\mbox {\bf m}}
\newcommand{\bn}{\mbox {\bf n}}
\newcommand{\bq}{\mbox {\bf q}}
\newcommand{\bx}{\mbox {\bf x}}
\newcommand{\bly}{\mbox {\bf y}}
\newcommand{\bz}{\mbox {\bf z}}
\newcommand{\br}{\mbox {\bf r}}
\newcommand{\bs}{\mbox {\bf s}}
\newcommand{\bt}{\mbox {\bf t}}
\newcommand{\bu}{\mbox {\bf u}}
\newcommand{\hbui}{\mbox {\bf \hat{u}_i}}
\newcommand{\hui}{\mbox {\hat{u}_i}}
\newcommand{\bw}{\mbox {\bf w}}
\newcommand{\bone}{\mbox {\bf 1}}
\newcommand{\bzer}{\mbox {\bf 0}}
\newcommand{\diag}{\mbox {diag}}
\newcommand{\var}{\mbox {var}}
\newcommand{\cov}{\mbox {cov}}
\newcommand{\cor}{\mbox {cor}}
\newcommand{\tr}{\mbox {tr}}
\newcommand{\ee}{\mbox {e}}
\newcommand{\lgt}{\mbox {logit}}
\newcommand{\beqn}{\begin {equation}}
\newcommand{\eeqn}{\end {equation}}
\newcommand{\beqa}{\begin {eqnarray}}
\newcommand{\eeqa}{\end {eqnarray}}
\newcommand{\ssqe}{\sigma^2_e}
\newcommand{\ssqu}{\sigma^2_u}
\newcommand{\half} {\frac {1}{2}}
\newcommand{\pqpt} {\partial {\bf q}/\partial {\bf \theta}}
\newcommand{\pqipt} {\partial {\bf q}_i/\partial {\bf \theta}}
\newcommand{\pqpts} {\partial^2 {\bf q}_i/\partial {\bf \theta} \partial {\bf \btheta^\top}}
\newcommand{\pfpt} {\partial {\bf h}/\partial {\bf \theta}}
\newcommand{\pfipt} {\partial {\bf h}_i/\partial {\bf \theta}}
\newcommand{\pfpts} {\partial^2 {\bf h}/\partial {\bf \theta} \partial {\bf \theta^\top}}
\newcommand{\pfipts}{\partial^2 {\bf h}_i/\partial {\bf \theta} \partial {\bf \theta^\top}}
\newcommand{\pfopts}{\partial^2 {\bf h}_1/\partial {\bf \theta} \partial {\bf \theta^\top}}
\newcommand{\pfmpts}{\partial^2 {\bf h}_m/\partial {\bf \theta} \partial {\bf \theta^\top}}
\newcommand{\pqipts}{\partial^2 {\bf q}_i/\partial {\bf \theta} \partial {\bf \theta^\top}}
\newcommand{\phpt}{\frac{\partial h}{\partial \btheta}}
\newcommand{\ft}{{\bf f(\theta,X_i)}}
\newcommand{\ftu}{{\bf f(\theta+u,X)}}
\newcommand{\fu} {{\bf f(\theta,X)}}
\newcommand{\sumi} {\sum_{i=1}^m}
\newcommand{\sumni} {\sum_{j=1}^{n_i}}
\newcommand{\ind} {\perp \!\!\!\perp}
\newcommand{\sigse} {\sigma_e^2}
\newcommand{\sigsu} {\sigma_u^2}
\newcommand{\sigu} {\sigma_u^2}
\newcommand{\pp}[2]{\frac{\partial #1}{\partial #2}}
\newcommand{\ppt}[2]{\frac{\partial^2 #1}{\partial #2 \partial {#2}^\top}}
\newcommand{\ppp}[3]{\frac{\partial^2 #1}{\partial #2 \partial #3}}
\newcommand{\ppl}[2]{\partial #1 / \partial #2}
\newcommand{\pppl}[3]{\partial^2 #1 / \partial #2 \partial #3 }
\newcommand{\bvbt}{\begin{verbatim}}
\newcommand{\benu}{\begin{enumerate}}
\newcommand{\eenu}{\end{enumerate}}
\newcommand{\bver}{\begin{verbatim}}
\def\beq{&~=~&}
\newcommand{\hbe}{\bar{\be}}
\newcommand{\nn}{\nonumber}
\makeatletter
\input{rotate}\newbox\rotbox
\newenvironment{sidetable}{\begin{table}[t]
\global\setbox\rotbox\vbox\bgroup
\hsize\textheight \@parboxrestore}%
{\par\vskip\z@\egroup \rotl\rotbox \end{table}}
\makeatother
\newdimen\x \x=1.5mm

\begin{abstract}
An accurate estimation of the dose-response relationship is important to determine the optimal dose. For this purpose, a dose finding trial in which subjects are randomized to a few fixed dose levels is the most commonly used design. Often, the estimation uses response data only, although drug exposure data are often obtained during the trial. The use of exposure data to improve this estimation is difficult, as exposure-response relationships are typically subject to confounding bias even in a randomized trial. We propose a robust approach to estimate the dose-response relationship without assuming a true exposure-response model, using dose as an instrumental variable. Our approach combines the control variable approach in causal inference with unobserved confounding factors and the ANCOVA adjustment of randomized trials.  The approach presented uses working models for dose-exposure-response data, but they are robust to model misspecification and remain consistent when the working models are far from correct. The asymptotic properties of the proposed approach are also examined.   A simulation study is performed to evaluate the performance of the proposed approach.  For illustration, the approach is used to a Car-T trial with randomized doses.

\end{abstract}

\section{Introduction}
To determine the optimal dose, comparisons of drug responses at different dose levels are fundamentally important \citep{fdadose}. For this purpose, a dose finding trial, in which subjects are randomized to a few fixed dose levels, is the most widely used trial design. The mean response at each dose level provides an unbiased estimate due to randomization. However, with small sample sizes in feasible dose-finding trials, there is often high variability in these estimates.  Drug exposure data such as AUC and trough concentration levels are also commonly collected; however, they are mostly used in (population) PK/PD modeling to provide information for dose determination from the clinical pharmacology aspect. Furthermore, although population PK / PD modeling is commonly used to predict response at a given dose, prediction may be very sensitive to misspecification of the dose-exposure (DE) model and the exposure-response (ER) model. 

To predict the outcome at a given dose level, the PK/PD modeling approach consists of two steps. 
 First, we fit both the DE and ER models to the exposure and response data separately.  For a given dose, we predict the exposure for individuals using the fitted DE model, then use the predicted exposure in the ER model to predict the response  \citep{wang2015exposure}.  The mean response can be estimated by averaging the predicted outcomes. In addition to sensitivity to model misspecification, one challenge is the potential confounding bias when fitting the ER model due to confounding factors that affect both the exposure and the response.   \cite{nedelman2007diagnostics} illustrated the importance of considering confounding in PK/PD modeling in a randomized dose trial of oxcarbazepine and approaches to diagnose it.
 
 To mitigate the impact of confounding bias without knowing the confounders, a powerful approach is to use randomized dose as instrumental variable (IV) \citep{wang2012dose, wang2014determining}.  Among early work using randomization as an IV in this area is \cite{nedelman2007diagnostics} who used IV in a specification test in PK/PD modeling for oxcarbazepine.  However, this approach is also model sensitive since, in general, the ER relationship is not nonparametrically identifiable with only a few fixed dose levels.  The issues of nonparametric and semiparametric identification using IVs have been discussed extensively in econometrics literature; See \cite{imbens2009identification}  and \cite{horowitz2011applied} for a review and \cite{zhang2025instrumental} for a recent application of IV methods in ER modeling.

We consider how to use exposure data to improve the estimation of DE relationship with minimum model assumption, in particular, for ER models. For this purpose, we propose an approach that combines the control variable (CV), also known as control function \citep{imbens2009identification} and analysis of covariance (ANCOVA) adjustment in RCTs, with g-computation including the CV in working models to improve the estimation of mean response at the doses in the trial. Our approach is closely related to ANCOVA adjustment for prognostic factors in RCTs for comparison between treatments \citep{Senn1989} and is similar to the adjustment for studies with pre-test post-test designs \citep{Leon_2003, Davidian_2005, Zhang_2008}. They showed that these adjustments are robust to model misspecification and can be semiparametric efficient.  For a finite population with the Neyman-Rubin framework, with only a few technical conditions, the ANCOVA-adjusted comparison is at least as good as the unadjusted group mean difference \citep{Lin_2013}. In this paper, we will focus on the former situation, as the latter requires a rather different approach.    

 The approach needs a key assumption on the DE model, as stated in \cite{imbens2009identification} for the construction of the CV. However, the approach requires almost no assumption on the ER model.  The approach does not estimate the ER relationship.  Instead, CV is used as a surrogate of potential unobserved confounding factors in a working ANCOVA model to improve the estimation of the dose-response (DR) relationship. Its improvement depends on how well the working model approximate the true one. However, the approach provides a consistent estimator even when the working models are very different from the true ones.  
 
The CV can be considered as a generalized propensity score \citep{Imai2004CausalIW}.  Although propensity scores are commonly used in observational studies, they can also be used in randomized trials to adjust for remaining imbalance of prognostic factors \citep{Williamson2013, Shen2013, Zeng2020}.  Therefore, our approach has a connection to the use of a generalized propensity score as a covariate for ANCOVA adjustment for RCTs.   However,  the major difference is that our approach adjusts for unobserved potential confounders.  With some specific models, the approach is also equivalent to the approach using PK measures as a covariate in dose-response modeling for dose-escalation trials \citep{piantadosi1996improved}, although no theoretical justification was given.

The next section introduces DE and ER models and provides a review of the CV approach with discrete IV. Our approach is described in Section 3, and its properties are examined in Section 4.  Section 5 presents results of a simulation study.  An application of the proposed approach to a Car-T trial is presented in Section 6.

\section{Dose-exposure-response models and relevant work}
In this section, we introduce DE and ER models and review relevant work in two areas: DE and ER modeling with IV-based CV approaches, and the ANCOVA adjustment for RCTs. This is not intended as a mini-review, but rather as a brief summary of works connected to our approach. 

\subsection{PK/PD modeling and causal inference}
We consider a simple situation of the ER relationship between exposure (through PK measures) $C_{ij}$  and response $Y_{ij}$ (for example, a glucose measure for a patient with type I diabetes) at the $j$th visit in a trial with randomized dose $D_i$ of subject $i$.  Suppose that $Y_{ij}, C_{ij}$ are measured at steady state, hence $E(Y_{ij}),E( C_{ij})$ do not change with $j$.  

\subsection{Dose-exposure-response relationship: triangular models}
The general DE-ER relationship can be presented by the following models
\begin{align}
    Y=& g(C,\epsilon)\\
    C=& h(D,\eta),
    \label{tri}
\end{align}
where $g(.,.)$ and $h(.,.)$ are unknown DE and ER models, $Y$ and $C$ are response and exposure, respectively, and $D$ is a randomized dose. $\epsilon$ and $\eta$ are potentially correlated random variables or vectors, since they may represent unobserved confounders $U$.  For example, $\epsilon$ and $\eta$ may represent age effect, which affect both $Y$ and $C$.   Covariates $X$ can be introduced in both models, but we will use the above simple form at this time. To determine the dose that yields the desired outcome, one approach is to identify both models $g(.,.)$ and $h(.,.)$.  Then, for any given $D$, one can predict the outcome $Y$ by $g(h(D,\eta),\epsilon)$.  This approach has been widely used in population PK / PD modeling, which fits both models including sufficient covariates,  then uses the fitted models to predict the outcome at individual and population levels \citep{wang2015exposure}.   The DE (for example, population PK) model can be easily fitted.  The major difficulty is to fit the ER model, due to the correlation between $\epsilon$ and $\eta$.  One may include sufficient covariates so that conditional on them,  $\epsilon$ and $\eta$ are independent. However, strong assumptions on sufficient available covariates and correct models are needed, yet the approach is still sensitive to model misspecification.  

The DE-ER relationship can be presented in a Directed Acyclic Graphs (DAG) (Figure \ref{dag}) in which $U$ represents unobserved confounders reflected in the correlation between $\epsilon$ and $\eta$.  Due to the existence of $U$, the ER relationship is subject to confounding bias and a simple analysis regressing $Y$ on $C$ is biased.  Note that the direct effect of $D$ on $Y$ is commonly excluded in the PK / PD modeling approach, while our proposed approach allows it.  
\begin{figure}
    \centering
    \includegraphics[width=100mm]{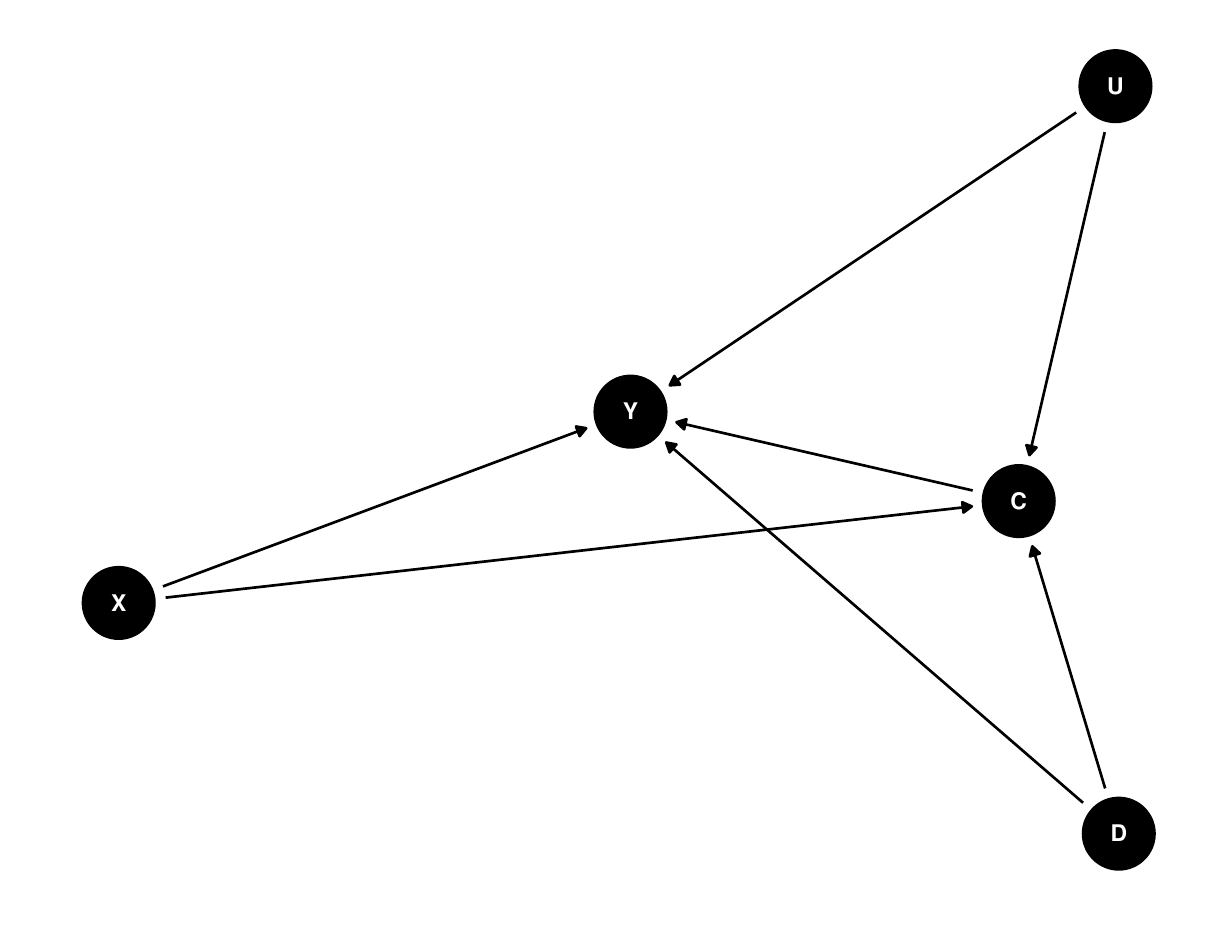}
    \caption{DAG for DER relationships with possible direct effect from $D$ to $Y$, observed covariates $X$ and unobserved covariates $U$ }
    \label{dag}
\end{figure}


In practice, only a few fixed doses $D=(d_1,...,d_K)$ can be tested in a trial. Therefore, a simpler task for dose finding is the comparison of the outcome between individual tested doses $d_j$ and $d_k$. Here, we can write the marginal dose-response relationship as 
\begin{equation}
E(Y(d))= \int_\epsilon \int_\eta g(h(d,\eta),\epsilon) dF_{\epsilon|\eta}(\epsilon) dF_\eta(\eta).  \label{ey}
\end{equation}
 Suppose that we want to estimate
\begin{equation}
\Delta_{jk}=E(Y(d_j)-E(Y(d_k)).    
\end{equation}
With $D$ randomized, an unbiased estimator of $\Delta_{jk}$ is
\begin{equation}
    \hat \Delta_{jk}^u=\bar y_j-\bar y_k,
\end{equation}
where $\bar y_j$ is the sample mean of $Y$s with dose $D=d_j$.  However, it does not use the information in $C$. Eq \ref{ey}  suggests that, if both $g(.,.)$ and $h(.,.)$ are known, $E(Y|d)$ can be calculated numerically.  In fact, this approach is equivalent to population PK / PD modeling which uses DE and ER models to predict the outcome at given dose levels for dose selection.  However, even with $D$ randomized,  $g(.,.)$ may not be identifiable, as discussed below. 

  \subsection{Control variables, discrete IV and identification of DE relationship}
As mentioned in the previous section, since the estimation of the DE relationship is easy,  the DR relationship can be obtained if we can identify the ER relationship.  The purpose of this section is to introduce the IV approach, particularly the use of CVs, and to show that, with only a few dose levels, identification of the ER relationship generally requires a parametric model assumption.

Suppose that the goal is to identify the marginal ER relationship 
\begin{equation}
 \mu(c)    \equiv \int g(c,\epsilon) dF_\epsilon(\epsilon).
\end{equation}
Due to unobserved confounder $U$ in Figure \ref{dag},  simply regressing $Y$ on $C$ is invalid.  Adjusting for $U$, the randomized dose as an IV is a powerful tool.  As $D$ is randomized, we can assume that

{\bf Assumption 1:}  $\epsilon, \eta \perp D$. 

Under this and additional assumptions, one may construct a control variable (or function) $V$ such that the following conditional independence holds
\begin{equation}
    C \perp \epsilon | V.
\end{equation}
Therefore, conditioning on $V$ can block the $U$ path to $Y$ in Figure \ref{dag} to eliminate its confounding effects.  Since $V$ is not directly observed, to condition on it, we need to construct a good estimate.
When the DE model is separable, i.e., it can be written as 
\begin{equation}
    C=h(D)+\eta,
\end{equation}
an easy way to construct an CV is to fit the model and then take the residuals
\begin{equation}
    V=C-\hat h(D),
\end{equation}
where $\hat h(D)$ is the fitted model.   In general, a CV can also be constructed under the following assumption:

{\bf Assumption 2:} $\eta$ is a continuously distributed scalar with CDF that is strictly increasing on its support and $h(D, t)$ is strictly monotonic in $t$ with probability one.

Under Assumptions 1 and 2,  \cite{imbens2009identification} showed that $V^*=F_{C|D} (\eta)$ is a CV (Theorem 1) for general including non-separable $h(D,\eta)$.  Although $V^*$ 
is not on the same scale as the residual CV, $ V=C-h(D)$, they are equivalent in the sense that since $F_{C|D} (\eta)$ is continuous and restrictively increasing,  conditioning on $V^*$ is equivalent to conditioning on $V$ for separable models. Note that $V^*=F_{C|D} (\eta)$ as CV can also be defined, conditioning on $D$ and covariates $X$, the same as the definition of generalized propensity score for general treatment in  \cite{Imai2004CausalIW}.  

To use $V$ in \eqref{muc}, a common approach is to make a model assumption such as 
\begin{equation}
    E(Y|C,V)= \beta_1 C +\beta_2 V,
\end{equation}
which is known as the residual inclusion approach \citep{wang2012dose, wang2014determining}.    However, this approach is sensitive to model misspecification.  As a general approach, we can define $m(c,v)=E(Y|C=c, V=v)$ as a function of both $c$ and $v$,  then $\mu(c)$ can be written as
\begin{equation}
    \mu(c)=\int m(c,v) dF_{V} (v)
    \label{muc}
\end{equation}
However, to identify $\mu(c)$ using the above formula, we also need that the support of $V$ is independent of $C$ (the assumption of common support of $V$, Assumption 2 in \cite{imbens2009identification}), which means that $V$ can be driven by $D$ to cover the whole support of $V$. As $V$ is continuous, hence its support is an interval.  As only a few fixed doses $(d_1,...,d_k)$ can be tested in the trial,  $D \in (d_1,...,d_k)$ is a discrete IV.  Hence, the common support assumption cannot be satisfied.  Therefore, $\mu(c)$ cannot be nonparametrically identified in our context without further assumptions on the models and $\epsilon$ and $\eta$. .

In summary, none of the classical approaches can identify the ER model without some parametric assumptions in our context.  Therefore, we turn to another task: to improve the estimation of the marginal DR model $\mu(d)=E(Y(d))$.    Again, full identification is not possible. Nevertheless, it can be identified at the tested dose levels, thanks to randomized dose levels.  In the rest of the paper, we show that CVs can be used to estimate the mean responses at tested dose levels $\mu_k \equiv \mu(d_k)$ more efficiently by using CVs as a pseudo covariates.  For this purpose, we use ANCOVA approaches for RCTs that will be briefly reviewed below.

\subsection{Covariate adjustment for treatment effect estimation for RCTs}
ANCOVA is a common approach to covariate adjustment for a more accurate estimation for treatment effects than group means. Most of the work in this area focuses on RCTs to compare an active treatment with a control treatment. However, these approaches can be adapted for our purpose by considering doses as individual treatments. The adjustment is based on a working model assuming 
\begin{equation}
    E(Y_i|X_i,D_i)=g(X_i,D_i,\beta),
    \label{wmodel}
\end{equation}
where $\beta$ is a vector of unknown parameters. The most commonly used are linear models with common coefficients for $X_i$:
\begin{equation}
    g(X_i,D_i,\beta)= (X_i,D_i)^T \beta
    \label{lin}
\end{equation}
After fitting the model and obtaining parameter estimates $\hat \beta$, an ANCOVA estimator for $\mu_k$ can be constructed as 
\begin{equation}
    \hat \mu_k=\bar y_k+ n^{-1}\sum_{i=1}^n g(X_i,d_k, \hat \beta)- n_k^{-1} \sum_{i \in S_k} g(X_i,d_k,\hat \beta),
    \label{hatmu3}
\end{equation}
where $S_k$ is the subset of $n_k$ subjects with $D_i=d_k$.  This estimator is equivalent to the doubly robust estimator for multi-value treatments in \cite{uysal2015doubly, Linden2015}. 
This estimator is a special case of the doubly robust estimator, which is simplified to our estimator  when $D_i$ is randomized. The estimator $\hat \mu_k$ using model \eqref{lin} is known as the ANCOVA I estimator, while the one using a model with an interaction term in \eqref{lin}, or equivalently using separate models for each dose level,  $ g(X_i,d_k,\beta)= X_i^T \beta_k$ for all $i \in S_k$,
is known as the ANCOVA II estimator.  Both estimators are consistent for $\mu_k$ even when the working model is wrong in an RCT. The ANCOVA I estimator may have a larger variance than that of $\bar y_k$, while an ANCOVA II  estimator is guaranteed to have no large variance  than $\bar y_k$, if the sample size is large enough.  This also holds in the finite population setting \citep{Lin_2013}.  However, since ANCOVA II models need $K$ times as many parameters as that of the ANCOVA I model, when the number of covariates is large, the small sample properties of ANCOVA II estimators may not necessarily be better than ANCOVA I.  In the following, we will use the ANCOVA II model, but generally the results also hold for the model \eqref{lin}.

The asymptotic properties of these estimators have been examined in many papers such as \cite{Davidian_2005, Leon_2003,Zhang_2008,bartlett2018covariate}.  In fact,  $g(X_i,D_i, \beta)$ can be an arbitrary function, but using $E(Y_i|X_i,D_i)=g(X_i,D_i,\beta)$ is optimal, and can achieve semi-parametric efficiency \citep{Leon_2003, bartlett2018covariate}.  However, the sandwich variance estimator for $\hat \mu_k$ \citep{white1980heteroskedasticity} is also valid for a wrongly specified model.  

When the working model \eqref{wmodel} is a generalized linear model with a canonical link function, such as a linear model or a logistic model with a logit link, $\hat \mu_k$  in \eqref{hatmu3}degenerates to 
\begin{equation}
    \hat \mu_k= n^{-1}\sum_{i=1}^n g(X_i,d_k, \hat \beta_k)
    \label{hatmu1}
\end{equation}
and its variance can be estimated by the sandwich estimator, which is conservative when the model is misspecified, but valid for both ANCOVA I and II estimators. 

\section{A robust adjustment for control variable for dose-response relationship}
We consider two situations: one assumes no DR model, while the other assumes a semi-parametric model with linear part for the dose. 
For the first situation, we combine the IV approach and the ANCOVA estimators by including the CV as a covariate in the ANCOVA model for a more accurate DR relationship than the unadjusted one.     
 To construct our estimator, we need the following assumption: 

{\bf Assumption 3:}   There exists a CV $V$ such that
\begin{equation}
    V \perp D |X
    \label{Vind}
\end{equation}
for any given $X$.

Randomized $D$ is not sufficient for this assumption, since, although in principle $F_{C|D,X}$ can be constructed as a CV, it may still depend on $D$.  See \cite{kasy2011identification} for further discussion.   A sufficient condition is a separable structure for the DE model. 
For example, this assumption holds with DE model 
\begin{equation}
C=(\gamma_0 +\gamma_x X) D +\eta    
\label{ERModel}
\end{equation}
  With this model, we have $V=C-(\gamma_0 +\gamma_x X) D=\eta$, and $\eta \perp D$.  In fact, the assumption holds with all separable models in the form of $C=h(D, X)+\eta$, with which $\eta$ can be estimated by the residuals after fitting this model.  
  
  Although the nonparametric construction of the CV proposed in  \cite{imbens2009identification} can construct CVs without the structural assumption for the DE model,  they may not satisfy Assumption 3.  Therefore, we assume that the DE model is separable and can be parameterised as 
\begin{equation}
    C_i=h(D_i, X_i, \gamma)+\eta_i,
    \label{demodel}
\end{equation}
assuming a known model $h(D_i, X_i, \gamma)$ with unknown parameters $\gamma$, but no structural assumption on the ER model.  It is not clear whether there exists a CV satisfying Assumption 3 for a more general form than the separable model.  However, the separable model seems sufficient to cover common PK models. 

With the above assumptions, we propose a two-stage approach that combines control variable and ANCOVA approaches. The algorithm in the following uses ANCOVA II with separate models for each dose.  They can be replaced by a single model with $\hat V_i$ and $D_i$ interaction, or an ANCOVA I model. 
\begin{enumerate}
    \item Fit the DE model to obtain $\hat \gamma$ and residuals $\hat V_i= C_i-h(D_i,X_i,\hat \gamma)$. 
    \item Fit a working DR model $E(Y_i|X_i, \hat V_i) =g( X_i, \hat V_i,\beta_k)$ for subjects $i \in S_k$: the subset of $i=1,...,n$ with $D_i=d_k$, and obtain $\hat \bbeta_k, k=1,...,K$.  This working model may be rather different from the true marginal DR model.
    \item For each $d_k$ of the $K$ dose levels, calculate 
    \begin{align}
        \hat \mu_k= & \sum_{i \in S_k} g(X_i, \hat V_i,\beta_k)/n_k;\\
        \hat \mu_k^a= & \sum_{i =1}^n g(X_i, \hat V_i,\beta_k)/n,
    \end{align}
    \item Construct a model adjusted estimator
    \begin{equation}
        \hat \mu_k^m=\bar y_k+ \hat \mu_k^a-\hat \mu_k.
        \label{mum}
    \end{equation}
\end{enumerate}
We have assumed that the effect of dose $d_k$ is fully represented by $\beta_k$. But, a dose-specific model  $g_k( X_i, \hat V_i,\beta_k)$ can also be used.  

Assumption 3 is key to robustness of $\hat \mu_k^m$ to misspecified model $g( X_i, V_i,\beta_k)$, as we need $E(g(X_i, V_i,\beta_k)|D_i=d_k)=E(g(X_i, V_i,\beta_k)$ such that $\hat \mu_k^a -\hat \mu_k \rightarrow 0$ with $n_k \rightarrow \infty$ even when the model is misspecified.  As we demonstrate in our application later, this assumption can be examined at least in principle.

To see why $\hat \mu_d^m$ is generally more accurate than $\bar y_d$, consider the simple situation where $C_i=\gamma D_i + \eta_i$ with an unknown prognostic factor $\eta_i$ as a random error.  Then we can fit this model to obtain the residuals $\hat V_i=C_i-\hat \gamma D_i \approx \eta_i$.  With a working model $g^*(V_i,  \beta_k)=\beta_k+V_i$, i.e. a model for dose level $d_k$ with $V_i$ as a covariate, the above algorithm is asymptotically equivalent to the ANCOVA II adjustment for $\eta_i$ \citep{Zhang_2008, Lin_2013}, which is known to be no worse than $\bar y_k$.    

In the special situation of only two dose levels and an ANOVA DE model, i.e., a model containing only group means, since the sum of residual of the DE model is zero, we have  $\sum_{i:D_i=1} \hat V_i +\sum_{i:D_i=0} \hat V_i=0$.  Consequently, when $n_0=n_1$, $\hat V_i$ is already balanced, hence adjusting for $\hat V_i$ will make no different in this situation.  

Next we consider using $\hat V_i$ to adjust for covariates in linear DR models so that, under some conditions, the adjustment may improve the efficiency but is robust to certain model misspecification.  We use the following working model
\begin{equation}
    Y_i=\beta_0+\beta_d D_i+\beta_v \hat V_i+\epsilon_i
    \label{lindr}
\end{equation}
  Our goal is to estimate $E(Y(d))$ as a linear function of $d$, adjusting  for $\hat V_i$.   Fitting this model to obtain the parameter estimates $\hat \beta_d$ and $\hat \beta_v$ is straightforward.  It is well known from standard linear model theory that when the model is correct,  adjusting for $V_i$ is beneficial.  In the next section we show that $\hat \beta_d$    is robust to model misspecification to some extent.

The working model can also be extended by replacing $\beta_d D_i$ in \eqref{lindr} with a nonlinear function $g_d(D_i, \beta_d)$, and  $\hat \beta_d$ still holds some robust properties.  It is also possible to use a nonlinear model such as a generalized linear model including $D_i$ and $V_i$ as covariates.  Then, the marginal DR model can be estimated by the g-computation.  However, the estimated DR model is no longer robust to model misspecification. 

\section{Asymptotic properties \label{asymp}}
Here we consider asymptotic properties of estimator $\hat \mu_d^m$ and show 

{\bf Proposition 1:} Under Assumptions 1-3 and some technical conditions, we have
\begin{enumerate}
    \item $\hat \mu_k^m \rightarrow \mu_k$ when $n_k \rightarrow \infty$ regardless of the working DR model.
    \item $\sqrt{n_k}(\hat \mu_k^m - \mu_k) \sim N(0,\sigma_k^2)$ with variance $\sigma_k^2$ depending on the model specification.

\end{enumerate}

These results are based on the ANCOVA II model, and asymptotic properties are in terms of sample size $n_k$.  When using an ANCOVA I model, the above also holds with $n_k$ replaced by $n$, but $\sigma_k^2$s may be higher if the model is not specified correctly.

To prove this proposition we show the following step-by-step:
\begin{enumerate}
    \item $\hat V_i \sim F_n(v) \rightarrow F(v)$, the true distribution function of $V_i$.
    \item  $\hat \beta_k - \beta^*_k =o(n_k^{-1/2})$, where $\beta^*_k$ are constant (See the Appendix), even when the working model is misspecified.
    \item  The central limit theorem holds for all components of $\hat \mu_k^m$, hence $\hat \mu_k^m$ is asymptotically normally distributed.

\end{enumerate}
At each step, we can borrow existing results with some adaptation from IV approaches \citep{imbens2009identification} or ANCOVA adjustment for RCTs \citep{Davidian_2005,Leon_2003,Lin_2013,bartlett2018covariate,Zhang_2008}, hence we will not provide a complete proof here. Technical details can be found in the Appendix, where for complex assumptions we refer the reader to  references on which our results are based. 

Next, we consider the asymptotic properties of the estimator $\hat \beta_d$ by fitting the model \eqref{lindr} by, e.g., an LS method. For convenience, we assume that $D_i, Y_i$ are centered, so there is no intercept in the model \eqref{lindr}. 
The following results seems rather different from proposition 1, but demonstrates a similar robustness of the adjustment with $\hat V_i$:

{\bf Proposition 2:} Under Assumptions 1-3 and some technical conditions, we have
\begin{enumerate}
    \item If the true model is
\begin{equation}
    Y_i =\beta_d^T D_i +g_v(V_i)+\epsilon_i
\end{equation}
with unknown $g_v(V)$,  then  we have  $\hat \beta_d \rightarrow \beta_d$ when $n \rightarrow \infty$.

\item If the true model is
\begin{equation}
    Y_i =g_d(D_i) +g_v(V_i)+\epsilon_i
\end{equation}
with unknown $g_d(D)$, then under some technical conditions we have  $\hat \beta_d \rightarrow \beta_d^*$, where \begin{equation}
    \beta^*_d =\mbox{argmin}_{\beta_d} E_d[(g_d(D_i)-\beta_d^T D_i)^2]
\end{equation} when $n \rightarrow \infty$.  That is,  $\hat \beta_d$ tends to the coefficients for the best linear approximation of $g_d(d)$, which depends on the randomization of dose only.
    \item $\sqrt{n}(\hat \beta_d - \beta_d ) \sim N(0,\sigma^2)$ with variance $\sigma^2$ depending on model specification in the first scenario.  The same holds for the second scenario with $\beta_d$ replaced by  $\beta_d^*$. 
\end{enumerate}

These properties  relate to the invariance of causal models in  \cite{Peters2016} in the sense that a causal dose-response relationship should not change due to adjustment for other factors and different adjustment methods/models.  

When the working model is nonlinear, the marginal DR model calculated by g-computation also has some good asymptotic properties when the model is correctly specified.  Since we are concentrating on robust estimators only,  we will not discuss further in this direction.

\section{A simulation study}
We conducted a simulation to compare the proposed estimator with the unadjusted one.  The following models and parameters are used.
\begin{itemize}
    \item Fixed dose levels $d_k=1,2,3$ with 1:1:1 random allocation to either 60 or 100 subjects.
    \item The data are generated by the following models
    \begin{align}
          C_i &=D_i+V_i, \:\:\: V_i \sim N(0,1)\\
          Y_i & =C_i+b_1 \exp(C_i+b_2 V_i)+0.5 V_i+U_i, \:\: U_i \sim N(0,1); \:\: \mbox{Normal data} \\
          Y_i & \sim Bin(p_i); \:\:\:  \mbox{Binary data}\nn\\
          p_i &=1/(1+\exp(0.5 -C_i-b_1 \exp(C_i+b_2 V_i)-0.5 V_i+U_i))
    \end{align}
    where for binary outcome simulation, a constant 0.5 is added to the linear predictor to avoid extreme $p_i$ values.  With this setting, $\mu_3$ is roughly 95\%.   The models are not separable when $b_1$ and $b_2$ are non-zero.
    \item The true mean response at each $d_k$ is estimated by a simulation sample of 100000 subjects.
    \item For each scenario, relative biases and variances are calculated using 5000 simulation runs. 
\end{itemize}
For analysis, we use three approaches: 1) An ANCOVA I DR model with $\hat V_i$ and $D_i$ as covariates, 2) An ANCOVA II DR model with $\hat V_i$ and $D_i$  as covariates for each dose level.    

For normally distributed $Y_i$, the relative biases and ratios of variances (adjusted to unadjusted) of the ANCOVA I, II estimators compared with those of the unadjusted estimator $\hat \mu^m$ at the three doses are presented in Table \ref{tnormal}. 
With the ANCOVA I estimator (first panel), although the working ER model is misspecified in all but one scenarios,  the adjusted estimator had comparable bias to those of unadjusted, while its variances were generally lower.   However, when there is severe model misspecification (rows 1 and 5), the variance of the adjusted one at dose level 1 may be higher than that of the unadjusted one.
The results of the ANCOVA II estimator is given in the second panel.  Its bias is slightly larger than the ANCOVA I estimate, but it has consistently lower variance ratios, except when there is no model misspecification, in which situation the two perform very similarly.   

The simulation result for binary $Y_i$ is presented in Table \ref{tbinary} for all three estimators.  A working logistic regression model is used for the DR and DE-ER models.  Relative biases are multiplied by 10, as they are much lower than those of the normally distributed $Y_i$.  An interesting finding here is that the ANCOVA I estimator has lower variances than the ANCOVA II estimator, which also did not show a variance reduction in $\hat \mu_3$.  

Finally,  as a scenario of severely misspecified models,  the simulation is repeated for binary $Y_i$ but using working linear models for DR and DE-ER models, and the simulation results are presented in Table \ref{tbinarylin} for all three estimators. Both estimators based on ANCOVA I and II have no efficiency gain for the $\mu_3$ estimator, and the former has up to 20\% variance inflation.  This is expected, as model misspecification has the highest impact at extreme probabilities, which occurs in $\mu_3$. Variance reduction can still be found for $\mu_1$ and $\mu_2$, but in general less than those with logistic models, showing the benefit of using a less misspecified model.

\begin{table}[ht]
\centering
\caption{Relative bias and variance of unadjusted and adjusted estimators responses based on ANCOVA I, II models at doses 1, 2 and 3, with varying $n, b_1$ and $b_2$ and 5000 simulation runs each scenario: normally distributed outcome. \label{tnormal}}
\begin{tabular}{rrrrrrrrrrrr}
  \hline
  &&& \multicolumn{3}{c}{Bias$\times$ 10 (unadjusted)} &\multicolumn{3}{c}{Bias$\times$ 10 (adjusted)} &\multicolumn{3}{c} {Var ratio (adj/unadj)}\\
 $n$ & $b_1$ & $b_2$ & $\mu_1$ & $\mu_2$ & $\mu_3$ & $\mu_1$ & $\mu_2$ & $\mu_3$ & $\mu_1$ & $\mu_2$ & $\mu_3$ \\ 
  \hline
 \multicolumn{12}{c}{ANCOVA I}\\
  60 & 0.3 & 0.2 & -0.02 & -0.02 & -0.02 & -0.00 & -0.01 & -0.03 & 1.42 & 0.68 & 0.76 \\ 
 60 & 0.1 & 0.2 & -0.02 & -0.01 & -0.02 & -0.01 & -0.01 & -0.02 & 0.60 & 0.58 & 0.73 \\ 
 60 & 0.3 & 0.0 & -0.02 & -0.01 & -0.01 & -0.01 & -0.01 & -0.02 & 0.99 & 0.58 & 0.73 \\ 
 60 & 0.0 & 0.0 & -0.02 & -0.01 & -0.00 & -0.01 & -0.01 & -0.00 & 0.39 & 0.54 & 0.76 \\ 
 100 & 0.3 & 0.2 & -0.01 & -0.01 & -0.01 & -0.01 & -0.01 & -0.02 & 1.53 & 0.68 & 0.81 \\ 
 100 & 0.1 & 0.2 & -0.01 & -0.01 & -0.01 & -0.01 & -0.00 & -0.01 & 0.61 & 0.57 & 0.78 \\ 
 100 & 0.3 & 0.0 & -0.01 & -0.01 & -0.01 & -0.01 & -0.00 & -0.01 & 0.97 & 0.57 & 0.77 \\ 
 100 & 0.0 & 0.0 & -0.01 & -0.00 & -0.00 & -0.01 & -0.00 & -0.00 & 0.39 & 0.53 & 0.79 \\ 
   \hline
   \multicolumn{12}{c}{ANCOVA II}\\
60 & 0.3 & 0.2 & -0.02 & -0.02 & -0.02 & -0.05 & -0.05 & -0.05 & 0.33 & 0.54 & 0.73 \\ 
 60 & 0.1 & 0.2 & -0.02 & -0.01 & -0.02 & -0.03 & -0.04 & -0.04 & 0.30 & 0.48 & 0.71 \\ 
 60 & 0.3 & 0.0 & -0.02 & -0.01 & -0.01 & -0.04 & -0.04 & -0.04 & 0.27 & 0.47 & 0.71 \\ 
 60 & 0.0 & 0.0 & -0.02 & -0.01 & -0.00 & -0.01 & -0.01 & -0.00 & 0.38 & 0.53 & 0.75 \\ 
 100 & 0.3 & 0.2 & -0.01 & -0.01 & -0.01 & -0.03 & -0.03 & -0.03 & 0.38 & 0.55 & 0.79 \\ 
 100 & 0.1 & 0.2 & -0.01 & -0.01 & -0.01 & -0.02 & -0.02 & -0.02 & 0.31 & 0.48 & 0.77 \\ 
 100 & 0.3 & 0.0 & -0.01 & -0.01 & -0.01 & -0.02 & -0.02 & -0.02 & 0.29 & 0.47 & 0.76 \\ 
 100 & 0.0 & 0.0 & -0.01 & -0.00 & -0.00 & -0.01 & -0.00 & -0.00 & 0.37 & 0.50 & 0.77 \\ 
 \hline
\end{tabular}
\end{table}

\begin{table}[ht]
\centering
\caption{Relative bias and variance of unadjusted and adjusted estimators responses based on ANCOVA I, II models at doses 1, 2 and 3, with varying $n, b_1$ and $b_2$ and 5000 simulation runs each scenario: binary outcome with logistic working model. \label{tbinary}}
\begin{tabular}{rrrrrrrrrrrr}
\hline 
  \hline
  &&& \multicolumn{3}{c}{Bias$\times$ 10 (unadjusted)} &\multicolumn{3}{c}{Bias$\times$ 10 (adjusted)} &\multicolumn{3}{c} {Var ratio (adj/unadj)}\\
 $n$ & $b_1$ & $b_2$ & $\mu_1$ & $\mu_2$ & $\mu_3$ & $\mu_1$ & $\mu_2$ & $\mu_3$ & $\mu_1$ & $\mu_2$ & $\mu_3$ \\ 
  \hline
   \multicolumn{12}{c}{ANCOVA I}\\
60 & 0.3 & 0.2 & 0.00 & -0.02 & 0.01 & -0.01 & -0.00 & -0.00 & 0.65 & 0.71 & 0.83 \\ 
 60 & 0.1 & 0.2 & -0.01 & -0.00 & -0.01 & -0.02 & 0.00 & -0.01 & 0.71 & 0.78 & 0.86 \\ 
 60 & 0.3 & 0.0 & -0.03 & 0.00 & 0.01 & -0.01 & 0.02 & -0.00 & 0.68 & 0.75 & 0.82 \\ 
 60 & 0.0 & 0.0 & -0.01 & -0.04 & -0.01 & -0.01 & -0.04 & -0.02 & 0.76 & 0.82 & 0.92 \\ 
 100 & 0.3 & 0.2 & -0.04 & -0.01 & 0.00 & -0.02 & -0.01 & -0.00 & 0.66 & 0.73 & 0.81 \\ 
 100 & 0.1 & 0.2 & -0.03 & -0.02 & 0.00 & -0.03 & -0.01 & -0.01 & 0.71 & 0.77 & 0.86 \\ 
 100 & 0.3 & 0.0 & -0.01 & 0.00 & -0.00 & 0.01 & 0.01 & -0.00 & 0.66 & 0.75 & 0.83 \\ 
 100 & 0.0 & 0.0 & 0.01 & 0.00 & -0.01 & 0.01 & -0.01 & -0.02 & 0.74 & 0.82 & 0.92 \\  
   \hline
   \multicolumn{12}{c}{ANCOVA II}\\
60 & 0.3 & 0.2 & 0.00 & -0.02 & 0.01 & -0.00 & 0.01 & 0.04 & 0.69 & 0.83 & 1.03 \\ 
60 & 0.1 & 0.2 & -0.01 & -0.00 & -0.01 & -0.01 & 0.00 & 0.00 & 0.74 & 0.87 & 1.03 \\ 
60 & 0.3 & 0.0 & -0.03 & 0.00 & 0.01 & -0.01 & 0.02 & 0.04 & 0.71 & 0.89 & 0.98 \\ 
60 & 0.0 & 0.0 & -0.01 & -0.04 & -0.01 & -0.01 & -0.05 & -0.02 & 0.79 & 0.88 & 1.01 \\ 
100 & 0.3 & 0.2 & -0.04 & -0.01 & 0.00 & -0.02 & -0.02 & 0.01 & 0.68 & 0.80 & 1.00 \\ 
100 & 0.1 & 0.2 & -0.03 & -0.02 & 0.00 & -0.03 & -0.02 & -0.01 & 0.72 & 0.81 & 0.98 \\ 
100 & 0.3 & 0.0 & -0.01 & 0.00 & -0.00 & 0.01 & 0.00 & 0.01 & 0.68 & 0.82 & 1.02 \\ 
100 & 0.0 & 0.0 & 0.01 & 0.00 & -0.01 & 0.02 & -0.01 & -0.02 & 0.76 & 0.86 & 0.96 \\ 
 \hline
\end{tabular}
\end{table}

\begin{table}[ht]
\centering
\caption{Relative bias and variance of unadjusted and adjusted estimators responses based on ANCOVA I, II models at doses 1, 2 and 3, with varying $n, b_1$ and $b_2$ and 5000 simulation runs each scenario: binary outcome with linear working models. \label{tbinarylin}}
\begin{tabular}{rrrrrrrrrrrr}
\hline 
  &&& \multicolumn{3}{c}{Bias$\times$ 10 (unadjusted)} &\multicolumn{3}{c}{Bias$\times$ 10 (adjusted)} &\multicolumn{3}{c} {Var ratio (adj/unadj)}\\
 $n$ & $b_1$ & $b_2$ & $\mu_1$ & $\mu_2$ & $\mu_3$ & $\mu_1$ & $\mu_2$ & $\mu_3$ & $\mu_1$ & $\mu_2$ & $\mu_3$ \\ 
  \hline
   \multicolumn{12}{c}{ANCOVA I}\\
60 & 0.3 & 0.2 & 0.00 & -0.02 & 0.01 & 0.01 & -0.01 & -0.00 & 0.74 & 0.77 & 1.16 \\ 
 60 & 0.1 & 0.2 & -0.01 & -0.00 & -0.01 & -0.01 & -0.00 & -0.01 & 0.77 & 0.81 & 1.05 \\ 
 60 & 0.3 & 0.0 & -0.03 & 0.00 & 0.01 & -0.01 & 0.01 & -0.01 & 0.77 & 0.82 & 1.23 \\ 
 60 & 0.0 & 0.0 & -0.01 & -0.04 & -0.01 & -0.01 & -0.04 & -0.01 & 0.79 & 0.84 & 1.00 \\ 
 100 & 0.3 & 0.2 & -0.04 & -0.01 & 0.00 & -0.02 & -0.01 & -0.01 & 0.76 & 0.81 & 1.16 \\ 
 100 & 0.1 & 0.2 & -0.03 & -0.02 & 0.00 & -0.03 & -0.01 & -0.00 & 0.77 & 0.81 & 1.04 \\ 
 100 & 0.3 & 0.0 & -0.01 & 0.00 & -0.00 & 0.00 & 0.00 & -0.01 & 0.76 & 0.81 & 1.21 \\ 
 100 & 0.0 & 0.0 & 0.01 & 0.00 & -0.01 & 0.01 & -0.01 & -0.01 & 0.78 & 0.84 & 0.99 \\  
   \hline
   \multicolumn{12}{c}{ANCOVA II}\\
 60 & 0.3 & 0.2 & 0.00 & -0.02 & 0.01 & 0.08 & 0.07 & 0.05 & 0.70 & 0.80 & 1.01 \\ 
 60 & 0.1 & 0.2 & -0.01 & -0.00 & -0.01 & 0.03 & 0.06 & 0.03 & 0.77 & 0.85 & 0.97 \\ 
 60 & 0.3 & 0.0 & -0.03 & 0.00 & 0.01 & 0.06 & 0.09 & 0.04 & 0.73 & 0.84 & 1.03 \\ 
 60 & 0.0 & 0.0 & -0.01 & -0.04 & -0.01 & 0.02 & 0.01 & 0.02 & 0.80 & 0.89 & 0.98 \\ 
 100 & 0.3 & 0.2 & -0.04 & -0.01 & 0.00 & 0.03 & 0.04 & 0.03 & 0.71 & 0.83 & 1.01 \\ 
 100 & 0.1 & 0.2 & -0.03 & -0.02 & 0.00 & -0.00 & 0.03 & 0.02 & 0.73 & 0.82 & 0.97 \\ 
 100 & 0.3 & 0.0 & -0.01 & 0.00 & -0.00 & 0.05 & 0.05 & 0.02 & 0.69 & 0.83 & 1.01 \\ 
 100 & 0.0 & 0.0 & 0.01 & 0.00 & -0.01 & 0.03 & 0.03 & 0.01 & 0.76 & 0.87 & 0.96 \\ 
   \hline 
\end{tabular}
\end{table}

\section{A numerical example}
We apply the proposed methods to a randomized dose optimization study of anti-CD19 chimeric
antigen receptor T (CART-19) cell therapy in patients with relapsed or refractory chronic
lymphocytic leukemia \citep{Frey2020}. The study compared two doses of CART-19 cell
therapy: a low dose (5×10$^7$ cells) and a high dose (5×$10^8$ cells). The study was conducted
in two stages: a dose-finding stage followed by an expansion stage. In the dose-finding
stage, 28 patients were randomized 1 : 1 to the two doses, and 24 were evaluable for tumor
response. Following an interim analysis, the study dropped the low dose and opened an
expansion cohort in which 10 more patients received the high dose and 8 were evaluable for
tumor response. Thus, a total of 32 patients were infused and evaluable for tumor response.
The proportion of patients with overall response (i.e., complete or partial response) was 31\%
(4/13) in the low dose group, and 53\% (10/19) in the high dose group. Following \cite{Frey2020}, we will combine the two stages and treat the whole study as randomized.
In this context, dose refers to the target number of CART-19 cells to be infused, and
exposure to the actual number of CART-19 cells infused, which may be substantially lower
than the target number (\cite{Frey2020}, Table 1). This notion of exposure differs in some
ways from familiar measures of drug exposure based on concentration, but is nonetheless
within the scope of the IV framework and the proposed IV method. In this application, we
define Y as an indicator for overall response (OR) and X as the base-10 logarithm of the actual
number of CART-19 cells infused.

For the DE model, we fitted a linear model to the log-infused dose with randomized dose levels as the only covariate.  Then we apply our approach to estimate the mean CR rate by randomized dose.  Since there are only two doses, the fitted model is equivalent to the mean log-infused dose by the randomized dose level. For the ER model we use logistic models for CR as the dependent variable and $V_i$ and randomized dose, with and without the interaction term for the ANCOVA II and ANCOVA I adjustments, respectively.  Note that with only two doses, the DE-ER model and ANCOVA I approaches are equivalent; hence the results of the former are not presented. Table \ref{carttab} shows the estimated mean OR rate with SE, and 96\% CI by randomized dose, with adjustment of ANCOVA I and II, compared to those of unadjusted.  The estimated mean CR rates with both adjustments are very similar to the unadjusted, as are the SE and 95\% CIs for the low dose. For high doses, the ANCOVA II estimate has a slightly lower SE and narrower 95\% CI than the ANCOVA I estimate, which is slightly better than the unadjusted one. Hence, our results show a slight benefit of using exposure data for adjustment.

As a sensitivity check, we plot the density of $\hat V_i$ by randomized dose in Figure \ref{fig1}.  The two curves are well aligned, although the low-dose ones show higher variability. In fact, this check is for illustration purpose, since, for only two doses, a significant imbalance is unlikely. Imbalance only occurs when a DE model is misspecified. 

\begin{table}[ht]
\centering
\caption{Estimated mean complete response rate with SE, and 96\% CI by randomized dose, with ANCOVA I and II adjustment, compared with those of unadjusted. \label{carttab}}
\begin{tabular}{rrrrrr}
  \hline
Method &Dose & Estimate & SE & lower CI & upper CI \\ 
  \hline
ANCOVA II & Low& 0.155 & 0.102 & -0.045 & 0.355 \\ 
   &High& 0.366 & 0.144 & 0.083 & 0.649 \\ 
\hline
ANCOVA I & Low&0.154 & 0.103 & -0.048 & 0.356 \\ 
   &High&  0.364 & 0.146 & 0.078 & 0.649 \\ 
\hline
No adjustment &Low& 0.154 & 0.102 & -0.047 & 0.354 \\ 
   &High& 0.364 & 0.148 & 0.073 & 0.654 \\ 
   \hline
\end{tabular}
\end{table}

\begin{figure}
    \centering
    \includegraphics[width=120mm]{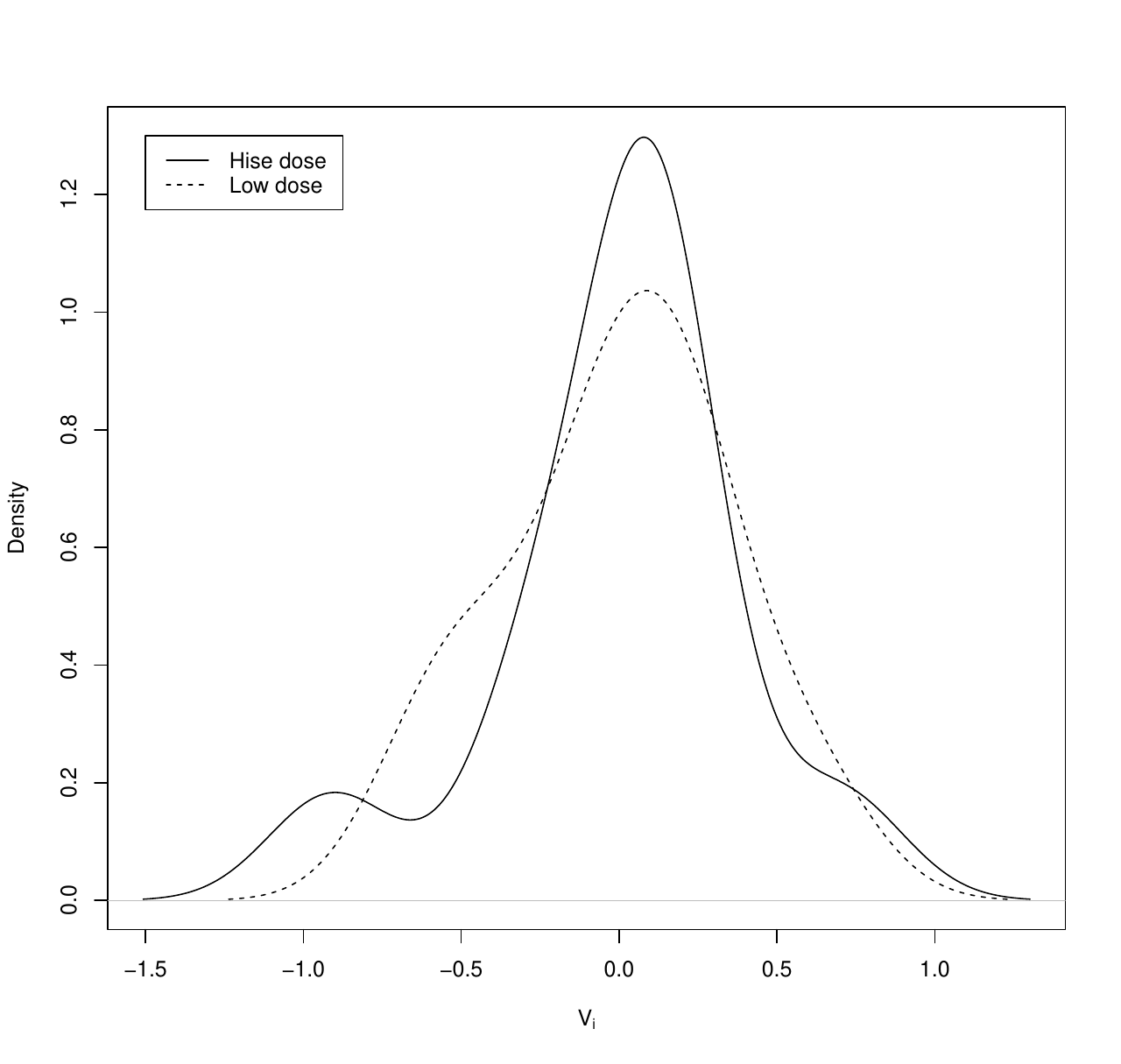}
    \caption{Distribution of $\hat V_i$ by dose level of Car-T data}
    \label{fig1}
\end{figure}

\section{Discussion}
Randomized dose comparison trials are widely used for dose selection.  The use of ANCOVA adjustment may improve the performance of the trials.  The exposure data from the trial are important information, as stated in the FDA guidance for oncology trials \citep{fdadose}, but they are not used for adjustment due to technical issues stated in the introduction. Although randomized dose can be used as an IV, identification of the ER relationship is still a challenging task. We propose a novel approach to use exposure data to generate a CV and use it as a covariate for the ANCOVA adjustment. The approach is model-agnostic in the sense that a simple working ANCOVA model such as a linear or logistic model can be used, and model misspecification does not affect consistency of the ANCOVA estimates. Furthermore, CV may partially capture the effect of unobserved prognostic factors and therefore may significantly improve ANCOVA estimators.  

Our approach relies on the assumption of independence between CV and dose, conditional on observed covariates.  In principle, this assumption can be verified by the trial data, although with small dose-finding trials, the verification may not be robust.  However, complete independence is not necessary, since it induces bias via the predicted model if the assumption is violated.  For a specific model, the assumption can be weakened according to a specific model.  For example, if a linear working model is used, the mean independence, i.e., $E(V|X,D)=E(V|X)$ is sufficient and easier to verify.  Checking this assumption in practice may still present challenges, hence we are not in a position to claim that this is an easy task.

Although randomized dose comparison trials are performed routinely, due to data confidentiality, a publishable data set is difficult to find.  For illustration,  we have applied our approach to the Car-T data \citep{Frey2020}.  The result shows only a very small gain in terms of lower SEs and narrower confidence intervals. As this is not a standard randomized dose trial, this result may not be representative. Using this approach on historical trial data will provide further insight on practical use and issues of the proposed approach.

We have focused on the estimation of $\mu_k$ only, since this task is sufficient to demonstrate our proposed approach.  Hypothesis tests for differences between dose levels, such as those implemented in MCPMod \citep{bretz2005combining} have been widely used for drug development. Based on the asymptotic properties of the proposed estimators, our approach can be used for hypothesis testing with or without multiplicity adjustment.  

One may also construct a working DR model as a combination of DE and ER models . Consider a linear ER model
\begin{equation}
    Y_i=\beta_c C_i +\beta_v V_i +\epsilon_i.
    \label{liny}
\end{equation}
With a dose-proportional DE model
\begin{equation}
    C_i=\gamma D_i + V_i,
\end{equation}
taking $C_i$ into the model \eqref{liny}, we have 
\begin{align}
    Y_i &= \beta_c \gamma D_i+ \beta_c v_i +\beta_v v_i  +\epsilon_i\nn\\
    &=\beta^*_d D_i +\beta^*_v v_i +\epsilon_i,  
\end{align}
where $\beta_d^* \equiv \beta_c \gamma$ and $\beta_v^* \equiv \beta_c +\beta_v $.
Therefore, the composite DE-ER model $g(,.,)$ is structurally equivalent to a linear empirical linear DR model.  However, the parameters can be estimated separately using the DE and ER data.  The invariance of the model structure generally does not hold for a nonlinear ER model.  For example, a logistic ER model and a linear DE model may not lead to a composite logistic model.  Nevertheless, our proposed approach is robust to model misspecification, hence the composite DE-ER model can also be used, even misspecified.   This approach is generally more complex than the ANCOVA I/II ones, and theoretical results of ANCOVA I/II cannot apply, hence we leave this topic to discuss separately.

We did not discuss the efficiency of the proposed approach, but some results for ANCOVA adjustment in the literature \citep{bartlett2018covariate, Zhang_2008} apply to our context.  These results show that $\hat \mu_k^m$ is semiparametric efficient, compared to all estimators adjusting for $\hat V_i$ only,  when the working models are correct.  This also holds when additional covariates $X_i$ are added to the model.  The benefit of adjustment depends on how much $\hat V_i$ can capture the impact of (unobserved) prognostic factors on $Y_i$.  

In designing of the trial, e.g., the number of dose levels and sample sizes are important for efficiency with or without CV adjustment.  The performance of the trial with our approach or other adjustments may be taken into account for sample size calculation in the design, although we acknowledge that it may be difficult to foresee the amount of improvement the proposed adjustment can make.

The use of exposure or PK data in randomized dose trials has been mainly the task for clinical pharmacologists.  The population PK/PD modeling approach has provided very useful information for dose selection.  Better use of PK data for this purpose requires collaboration between teams with different scientific expertise, including statisticians.  We expect further development that will lead to a better selection of doses for new drugs on the market.

\bibliographystyle{plainnat}
\bibliography{refs}

\begin{thebibliography}{27}
\providecommand{\natexlab}[1]{#1}
\providecommand{\url}[1]{\texttt{#1}}
\expandafter\ifx\csname urlstyle\endcsname\relax
  \providecommand{\doi}[1]{doi: #1}\else
  \providecommand{\doi}{doi: \begingroup \urlstyle{rm}\Url}\fi

\bibitem[Bartlett(2018)]{bartlett2018covariate}
Jonathan~W Bartlett.
\newblock Covariate adjustment and estimation of mean response in randomised trials.
\newblock \emph{Pharmaceutical statistics}, 17\penalty0 (5):\penalty0 648--666, 2018.

\bibitem[Bretz et~al.(2005)Bretz, Pinheiro, and Branson]{bretz2005combining}
Frank Bretz, Jos{\'e}~C Pinheiro, and Michael Branson.
\newblock Combining multiple comparisons and modeling techniques in dose-response studies.
\newblock \emph{Biometrics}, 61\penalty0 (3):\penalty0 738--748, 2005.

\bibitem[Davidian et~al.(2005)Davidian, Tsiatis, and Leon]{Davidian_2005}
Marie Davidian, Anastasios~A. Tsiatis, and Selene Leon.
\newblock Semiparametric estimation of treatment effect in a pretest{\textendash}posttest study with missing data.
\newblock \emph{Statistical Science}, 20\penalty0 (3), aug 2005.
\newblock \doi{10.1214/088342305000000151}.
\newblock URL \url{https://doi.org/10.1214%2F088342305000000151}.

\bibitem[Food and Administration(2023)]{fdadose}
Food and Drug Administration.
\newblock Optimizing the dosage of human prescription drugs and biological products for the treatment of oncologic diseases guidance for industry.
\newblock 2023.
\newblock URL \url{https://www.fda.gov/media/164555/download}.

\bibitem[Frey et~al.(2020)Frey, Gill, Hexner, Schuster, Nasta, Loren, Svoboda, Stadtmauer, Landsburg, Mato, Levine, Lacey, Melenhorst, Veloso, Gaymon, Pequignot, Shan, Hwang, June, and Porter]{Frey2020}
Noelle~V. Frey, Saar Gill, Elizabeth~O. Hexner, Stephen Schuster, Sunita Nasta, Alison Loren, Jakub Svoboda, Edward Stadtmauer, Daniel~J. Landsburg, Anthony Mato, Bruce~L. Levine, Simon~F. Lacey, Jan~Joseph Melenhorst, Elizabeth Veloso, Avery Gaymon, Edward Pequignot, Xinhe Shan, Wei-Ting Hwang, Carl~H. June, and David~L. Porter.
\newblock Long-term outcomes from a randomized dose optimization study of chimeric antigen receptor modified t cells in relapsed chronic lymphocytic leukemia.
\newblock \emph{Journal of Clinical Oncology}, 38\penalty0 (25):\penalty0 2862--2871, September 2020.
\newblock \doi{10.1200/jco.19.03237}.
\newblock URL \url{https://doi.org/10.1200/jco.19.03237}.

\bibitem[Horowitz(2011)]{horowitz2011applied}
Joel~L Horowitz.
\newblock Applied nonparametric instrumental variables estimation.
\newblock \emph{Econometrica}, 79\penalty0 (2):\penalty0 347--394, 2011.

\bibitem[Imai and van Dyk(2004)]{Imai2004CausalIW}
Kosuke Imai and David~A. van Dyk.
\newblock Causal inference with general treatment regimes.
\newblock \emph{Journal of the American Statistical Association}, 99:\penalty0 854 -- 866, 2004.
\newblock URL \url{https://api.semanticscholar.org/CorpusID:7489892}.

\bibitem[Imbens and Newey(2009)]{imbens2009identification}
Guido~W Imbens and Whitney~K Newey.
\newblock Identification and estimation of triangular simultaneous equations models without additivity.
\newblock \emph{Econometrica}, 77\penalty0 (5):\penalty0 1481--1512, 2009.

\bibitem[Kasy(2011)]{kasy2011identification}
Maximilian Kasy.
\newblock Identification in triangular systems using control functions.
\newblock \emph{Econometric Theory}, 27\penalty0 (3):\penalty0 663--671, 2011.

\bibitem[Leon et~al.(2003)Leon, Tsiatis, and Davidian]{Leon_2003}
Selene Leon, Anastasios~A. Tsiatis, and Marie Davidian.
\newblock Semiparametric estimation of treatment effect in a pretest-posttest study.
\newblock \emph{Biometrics}, 59\penalty0 (4):\penalty0 1046--1055, dec 2003.
\newblock \doi{10.1111/j.0006-341x.2003.00120.x}.
\newblock URL \url{https://doi.org/10.1111%2Fj.0006-341x.2003.00120.x}.

\bibitem[Lin(2013)]{Lin_2013}
Winston Lin.
\newblock Agnostic notes on regression adjustments to experimental data: Reexamining freedman's critique.
\newblock \emph{The Annals of Applied Statistics}, 7\penalty0 (1), mar 2013.
\newblock \doi{10.1214/12-aoas583}.
\newblock URL \url{https://doi.org/10.1214%2F12-aoas583}.

\bibitem[Linden et~al.(2015)Linden, Uysal, Ryan, and Adams]{Linden2015}
Ariel Linden, S.~Derya Uysal, Andrew Ryan, and John~L. Adams.
\newblock Estimating causal effects for multivalued treatments: a comparison of approaches.
\newblock \emph{Statistics in Medicine}, 35\penalty0 (4):\penalty0 534--552, October 2015.
\newblock \doi{10.1002/sim.6768}.
\newblock URL \url{https://doi.org/10.1002/sim.6768}.

\bibitem[Nedelman et~al.(2007)Nedelman, Rubin, and Sheiner]{nedelman2007diagnostics}
Jerry~R Nedelman, Donald~B Rubin, and Lewis~B Sheiner.
\newblock Diagnostics for confounding in pk/pd models for oxcarbazepine.
\newblock \emph{Statistics in medicine}, 26\penalty0 (2):\penalty0 290--308, 2007.

\bibitem[Peters et~al.(2016)Peters, B\"{u}hlmann, and Meinshausen]{Peters2016}
Jonas Peters, Peter B\"{u}hlmann, and Nicolai Meinshausen.
\newblock Causal inference by using invariant prediction: Identification and confidence intervals.
\newblock \emph{Journal of the Royal Statistical Society Series B: Statistical Methodology}, 78\penalty0 (5):\penalty0 947--1012, October 2016.
\newblock \doi{10.1111/rssb.12167}.
\newblock URL \url{https://doi.org/10.1111/rssb.12167}.

\bibitem[Piantadosi and Liu(1996)]{piantadosi1996improved}
Steven Piantadosi and Guanghan Liu.
\newblock Improved designs for dose escalation studies using pharmacokinetic measurements.
\newblock \emph{Statistics in medicine}, 15\penalty0 (15):\penalty0 1605--1618, 1996.

\bibitem[Senn(1989)]{Senn1989}
S.~J. Senn.
\newblock Covariate imbalance and random allocation in clinical trials.
\newblock \emph{Statistics in Medicine}, 8\penalty0 (4):\penalty0 467--475, April 1989.
\newblock \doi{10.1002/sim.4780080410}.
\newblock URL \url{https://doi.org/10.1002/sim.4780080410}.

\bibitem[Shen et~al.(2013)Shen, Li, and Li]{Shen2013}
Changyu Shen, Xiaochun Li, and Lingling Li.
\newblock Inverse probability weighting for covariate adjustment in randomized studies.
\newblock \emph{Statistics in Medicine}, 33\penalty0 (4):\penalty0 555--568, September 2013.
\newblock \doi{10.1002/sim.5969}.
\newblock URL \url{https://doi.org/10.1002/sim.5969}.

\bibitem[Uysal(2015)]{uysal2015doubly}
S~Derya Uysal.
\newblock Doubly robust estimation of causal effects with multivalued treatments: an application to the returns to schooling.
\newblock \emph{Journal of Applied Econometrics}, 30\penalty0 (5):\penalty0 763--786, 2015.

\bibitem[Van~der Vaart(2000)]{van2000asymptotic}
Aad~W Van~der Vaart.
\newblock \emph{Asymptotic statistics}, volume~3.
\newblock Cambridge university press, 2000.

\bibitem[Wang(2012)]{wang2012dose}
Jixian Wang.
\newblock Dose as instrumental variable in exposure--safety analysis using count models.
\newblock \emph{Journal of Biopharmaceutical Statistics}, 22\penalty0 (3):\penalty0 565--581, 2012.

\bibitem[Wang(2014)]{wang2014determining}
Jixian Wang.
\newblock Determining causal exposure-response relationships with randomized concentration-controlled trials.
\newblock \emph{Journal of Biopharmaceutical Statistics}, 24\penalty0 (4):\penalty0 874--892, 2014.

\bibitem[Wang(2015)]{wang2015exposure}
Jixian Wang.
\newblock \emph{Exposure-Response Modeling: Methods and Practical Implementation}, volume~84.
\newblock CRC press, 2015.

\bibitem[White(1980)]{white1980heteroskedasticity}
Halbert White.
\newblock A heteroskedasticity-consistent covariance matrix estimator and a direct test for heteroskedasticity.
\newblock \emph{Econometrica: journal of the Econometric Society}, pages 817--838, 1980.

\bibitem[Williamson et~al.(2013)Williamson, Forbes, and White]{Williamson2013}
Elizabeth~J. Williamson, Andrew Forbes, and Ian~R. White.
\newblock Variance reduction in randomised trials by inverse probability weighting using the propensity score.
\newblock \emph{Statistics in Medicine}, 33\penalty0 (5):\penalty0 721--737, September 2013.
\newblock \doi{10.1002/sim.5991}.
\newblock URL \url{https://doi.org/10.1002/sim.5991}.

\bibitem[Zeng et~al.(2020)Zeng, Li, Wang, and Li]{Zeng2020}
Shuxi Zeng, Fan Li, Rui Wang, and Fan Li.
\newblock Propensity score weighting for covariate adjustment in randomized clinical trials.
\newblock \emph{Statistics in Medicine}, 40\penalty0 (4):\penalty0 842--858, November 2020.
\newblock \doi{10.1002/sim.8805}.
\newblock URL \url{https://doi.org/10.1002/sim.8805}.

\bibitem[Zhang et~al.(2008)Zhang, Tsiatis, and Davidian]{Zhang_2008}
Min Zhang, Anastasios~A. Tsiatis, and Marie Davidian.
\newblock Improving efficiency of inferences in randomized clinical trials using auxiliary covariates.
\newblock \emph{Biometrics}, 64\penalty0 (3):\penalty0 707--715, jan 2008.
\newblock \doi{10.1111/j.1541-0420.2007.00976.x}.
\newblock URL \url{https://doi.org/10.1111%2Fj.1541-0420.2007.00976.x}.

\bibitem[Zhang et~al.(2025)Zhang, Wang, and Xi]{zhang2025instrumental}
Zhiwei Zhang, Jixian Wang, and Dong Xi.
\newblock An instrumental variable approach to learning the causal exposure--response relationship in a randomized dose comparison trial.
\newblock \emph{Statistics in Biopharmaceutical Research}, 17\penalty0 (1):\penalty0 125--135, 2025.

\end{thebibliography}

\section{Appendix}
\subsection{Supplementary to Section \ref{asymp}  }
Here, we provide more details of the proof for the asymptotic properties given in Propositions 1 and 2 in Section \ref{asymp}, using existing results and refer to the corresponding publication for technical conditions rather than repeating them here.  For the first of the three steps, we borrow the results from \cite{imbens2009identification}. It showed that in the general situation under Assumptions 1 and 2,  even for non-separable DE model, one can construct $\hat V_i$  such that $\max|\hat V_i -V_i| =O(n^{-1/2})$.  This result also holds for separable DE models.    
We assume that 
\begin{equation}
    E(C_i|D_i,X_i, v_i) =h(D_i, X_i, \gamma^0)+v_i
\end{equation}
which is correctly specified and $\gamma^0$ is the true parameter value.   
With this model, we can estimate $\gamma$ with an estimating equation (EE)
\begin{equation}
    0=\sum_{i=1}^n H_c(D_i, X_i) (C_i-h(D_i, X_i,  \gamma))/n    
    \label{eegamma}
\end{equation}
to obtain $\hat \gamma$.  Under mild technical conditions, we have $\hat \gamma - \gamma^0= O(n^{-1/2})$. 
Then, we can obtain $\hat V_i=C_i-h(D_i,\hat \gamma)=V_i +(h(D_i, \gamma)-h(D_i,\hat \gamma))=V_i +O(n^{-1/2})$,
where the last equation uses the assumption that  $h(D_i,\eta)$ is a continuous function of $\eta$.
Therefore, we have $\hat V_i \sim F_n(v)$ and $F_n(v) \rightarrow F(v)$: the true distribution function for $V$.  

Next, we assume that $\hat \beta_k$  is the solution to the following EE
\begin{equation}
    P_{n_k} e(X_i,Y_i, \hat V_i,\beta_k) \equiv n_k^{-1} \sum_{i \in S_k} e(X_i,Y_i,\hat V_i,\beta_k)=0,
\end{equation}
where $e(X_i,Y_i,\hat V_i,\beta_k)=G(X_i,\hat V_i,\beta_k)(Y_i-g( X_i,  \hat V_i,\beta_k))$, and $$G(X_i,\hat V_i,\beta_k)=\partial g(X_i, \hat V_i,\beta_k)/ \partial \beta_k$$ for maximum efficiency when $g( X_i,  \hat V_i,\beta_k)$
is a correct model for $E(Y_i|X_i,\hat V_i)$ for $i \in S_k$, but can be another function.  We show that when $E(e(X_i,Y_i,V_i,\beta_k)=0$ has a unique solution $\beta^*_k$,  under conditions for Lemma 5.1 of \cite{van2000asymptotic} , we have $\hat \beta_k \rightarrow \beta_k^*$ in probability.  To this end, we assume that $E(e(X_i,Y_i,\hat V_i,\beta_k))=\delta_k$ exists and $E(e(X_i,Y_i, h(D_i, X_i,  \gamma) ,\beta_k))$ has a finite derivative with respect to $\gamma$ to allow the order change between taking expectation and derivative.  We also assume that $\hat \gamma$ is the mean of influence functions of the EE \eqref{eegamma}, hence $\cov(\hat \gamma, e(X_i,Y_i,\hat V_i,\beta_k))= O_p(n^{-1/2} )$.  With Taylor expansion around $\gamma$ 
\begin{align}
      E(e(X_i,Y_i,\hat V_i,\beta_k)-e(X_i,Y_i,V_i,\beta_k))
    &= E(\frac{\partial e}{\partial \gamma}) E(\gamma-\hat \gamma) + O(n^{-1/2} )\nn\\
    &=O(n^{-1/2} )
\end{align}
where the second equality is due to that $E(\gamma-\hat \gamma) = O_p(n^{-1/2} )$ and $E(\partial e(X_i,Y_i,h(D_i, X_i,  \gamma),\beta_k)/\partial \gamma)$ is bounded.
Then, we have
\begin{align}
     P_{n_k} e(X_i,Y_i,\hat V_i,\beta_k) &=P_{n_k} e(X_i,Y_i,\hat V_i,\beta_k)+E(e(X_i,Y_i, V_i,\beta_k)) \nn\\
     &=P_{n_k} e(X_i,Y_i,\hat V_i,\beta_k)-E(e(X_i,Y_i,\hat V_i,\beta_k))\nn\\
     & +E(e(X_i,Y_i,\hat V_i,\beta)-e(X_i,Y_i, V_i,\beta_k))\nn\\
     &=O_p(n_k^{-1/2})
     \end{align}
By Lemma 5.10 in \cite{van2000asymptotic}, we have $\hat \beta_k \rightarrow \beta_k^*$ in probability when $n_k \rightarrow \infty$. 
In addition, if some technical conditions in Theorem 5.21 in \cite{van2000asymptotic} are satisfied by $e(X_i,Y_i,V_i,\beta_k)$, then $\sqrt{n_k} (\hat \beta_k -\beta_k^*)$ has a normal distribution with zero mean.  

Next, we show the central limit theorem for $\hat \mu_k^m$:
\begin{equation}
    \sqrt{n_k} (\hat \mu_k^m-\mu_k) \sim N(0,\sigma^2_k)
    \label{CLT}
\end{equation}
when $n_k \rightarrow \infty$. It is sufficient to show that $\bar y_k$, $ \hat \mu_k^a$and $\hat \mu_k $ all follow a normal distribution.  It is trivial for the first one.  For the other two, we assume that the working model $g(X_i, V_i,\beta_k)$ is continuous with respect to $V_i$ and $\beta$, and satisfies  conditions for a CLT such as the  Lindeberg–Lévy conditions on $E(g(X_i, V_i,\beta_k))=g_k$ and $\var(g( X_i, V_i,\beta_k))=\Sigma_k$  \citep{van2000asymptotic} .  Therefore, we have asymptotically
\begin{equation}
    \hat \mu_k=n_k^{-1} \sum_{i \in S_k} g(X_i, V_i,\beta_k) \sim N(g_k,\Sigma_k)
\end{equation}
Since $\hat V_i-V_i=O(n^{-1/2})$ and $\hat \beta_k-\beta=O(n_k^{-1/2})$, and $g(D_i, X_i, V_i,\beta_k)$ is continuous with respect to $V_i$ and $\beta$, $n_k^{-1} \sum_{i \in S_k} g( X_i, \hat V_i,\hat \beta_k)$ follows the same asymptotic distribution.  The same can be shown for $\hat \mu_k^a$ and finally we have shown \eqref{CLT}. 

The proof of proposition 2 is very similar to that of proposition 1.  Fitting the DR model can also be considered as solving an EE 
\begin{equation}
    P_{n} e(X_i,Y_i, \hat V_i,\beta) \equiv n^{-1} \sum_{i=1}^n e(X_i,Y_i,\hat V_i,\beta)=0,
    \label{linee}
\end{equation}
where
\begin{align}
    e(X_i,Y_i,\hat V_i,\beta)=&G(X_i,\hat V_i,\beta)(Y_i-g( X_i,  \hat V_i,\beta))\nn\\
    =& (D_i,\hat V_i)^T(Y_i-\beta_d D_i- \beta_v \hat V_i)
\end{align}
 The robustness of $\hat \beta_d$ is based on the expectation of $ P_{n} e(X_i,Y_i, \hat V_i,\beta)$ under a misspecified working model. 
 The weak convergence of $\hat V_i$ to $V_i$ is independent of the DR model, hence we have $P_{n} e(X_i,Y_i, \hat V_i,\beta)  \rightarrow P_{n} e(X_i,Y_i, V_i,\beta) $ weakly, and hence we can replace $\hat V_i$ with $V_i$.    
 
When the true model is $Y_i=\beta_d D_i+g_v(V_i)+\epsilon_i$, the component of $e(X_i,Y_i,\hat V_i,\beta)$ for $\beta_d$ has the expectation
\begin{align}
     E[D_i(Y_i-\beta_d D_i- \beta_v \hat V_i)]=& E[D_i(g_v(V_i)+\epsilon_i- \beta_v \hat V_i)]\nn\\
     =&0
\end{align}
 where the second equation is due to that $E(D_i)=0$ and $D_i$ and $V_i$ are independent.  Therefore, as a solution to EE \eqref{linee}, $\hat \beta_d \rightarrow \beta_d$ as $n \rightarrow \infty$.   Note that $\hat \beta_v \rightarrow \beta^*_v$, the solution to $ E[V_i(g_v(V_i)+\epsilon_i- \beta_v \hat V_i)]=0$.  

 When the true model is $Y_i=g_d(D_i)+g_v(V_i)+\epsilon_i$,  the component of $e(X_i,Y_i,\hat V_i,\beta)$  for $\beta_d$ has expectation
\begin{align}
     & E[D_i(Y_i-\beta_d D_i- \beta_v \hat V_i)]\nn\\
     =& E[D_i(g_d(D_i)-\beta_d D_i+ g_v(V_i)+\epsilon_i- \beta_v \hat V_i)]\nn\\
     =& E[D_i(g_d(D_i)-\beta_d D_i)]
\end{align}
 where the second equation is due to that $E(D_i)=0$ and $D_i$ and $V_i$ are independent.  Therefore, as a solution to EE \eqref{linee}, $\hat \beta_d \rightarrow \beta_d^*$ the solution to $  E[D_i(g_d(D_i)-\beta_d D_i)]=0$,  as $n \rightarrow \infty$.  

 The CLT for $\hat \beta_d$ (and $\hat \beta_v$) follows the standard results of estimators of misspecified models, such as \cite{white1980heteroskedasticity}.

\end{document}